\RequirePackage{commath, amsmath, amssymb, amsfonts}
\documentclass[10pt, a4paper]{iopart}

\usepackage{xcolor}
\usepackage{physics}
\usepackage{graphicx, float}
\usepackage{amsmath}
\usepackage{tikz}
\usepackage{hyperref}
\usepackage{hyperref}
\hypersetup{colorlinks,allcolors=black}
\usepackage{physics}
\usepackage{amsmath}
\usepackage{mathdots}
\usepackage{yhmath}
\usepackage{cancel}
\usepackage{color}
\usepackage{siunitx}
\usepackage{array}
\usepackage{multirow}
\usepackage{amssymb}
\usepackage{gensymb}
\usepackage{tabularx}
\usepackage{booktabs}
\usepackage{ulem}
\usepackage{nicematrix}
\usepackage{arydshln}

\begin{document}

\tikzset{every picture/.style={line width=0.75pt}}
\title{High-throughput search for triplet point defects with narrow emission lines in 2D materials}
\author{A. Sajid$^{\ast, 1,2}$, F. Nilsson$^{1}$, S. Manti$^{1,3}$, F. Bertoldo$^{1}$, J. J. Mortensen$^{1}$, K. S. Thygesen$^{1}$}
\address{$^1$CAMD, Computational Atomic-Scale Materials Design, Department of Physics, Technical University of Denmark, 2800 Kgs. Lyngby Denmark}
\address{$^2$School of Physics and Astronomy, Monash University, Victoria 3800, Australia}
\address{$^3$INFN, Laboratori Nazionali di Frascati, Via E. Fermi 54, I-00044 Roma, Italy;}

\address{$^{\ast}$Corresponding Author: sajal@dtu.dk}

\vspace{10pt}

\begin{abstract}
 We employ a first-principles computational workflow to screen for optically accessible, high-spin point defects in wide band gap two-dimensional (2D) crystals. Starting from an initial set of 5388 point defects, comprising both intrinsic and extrinsic, single and double defects in ten previously synthesised 2D host materials, we identify 596 defects with a triplet ground state. For these defects, we calculate the defect formation energy, the hyperfine (HF) coupling, and the zero-field splitting (ZFS) tensors. For 39 triplet transitions exhibiting particularly low Huang-Rhys factors, we calculate the full photo-luminescence (PL) spectrum. Our approach reveals many new spin defects with narrow PL line shapes and emission frequencies covering a broad spectral range. Most of the defects are hosted in hexagonal BN, which we ascribe to its high stiffness, but some are also found in MgI$_{2}$, MoS$_2$, MgBr$_2$ and CaI$_2$. As specific examples, we propose the defects $\mathrm{v_SMo_S}^{0}$ and $\mathrm{Ni_SMo_{S}}^{0}$ in MoS$_2$ as interesting candidates with potential applications to magnetic field sensors and quantum information technology. All the data will be made available in the open access database QPOD.   

\end{abstract}

\maketitle
\ioptwocol

\section{Introduction}\label{sec:introduction}
Two-dimensional (2D) materials hosting crystal point defects constitute an interesting platform for controlling electron spins with light. In particular, optically detected magnetic resonance (ODMR) can be used to initialise and read the state of a defect spin by means of optical pumping and fluorescence measurement. The ODMR technique has been extensively applied to the NV$^-$ defect in diamond targeting a range of applications such as optically addressed magnetic sensors\cite{gruber1997scanning,taylor2008high,chipaux2015magnetic}, quantum light sources operating at room temperature\cite{sipahigil2012quantum}, biomedical imaging\cite{zhang2021toward}, and quantum information processing\cite{weber2010quantum}. Recently, point defects in the 2D material hexagonal boron nitride (hBN)\cite{chejanovsky2021single,stern2022room,gottscholl2020initialization} have also been shown to exhibit ODMR effects.  

While defects buried inside a bulk crystal often have good structural and chemical stability thanks to the rigid solid state environment, they can be difficult to create and probe with high spatial precision. The situation is quite the opposite for defects in 2D materials where the inherent surface proximity render the defects easier to address and manipulate but also render them less stable and more sensitive to the surrounding environment. The stability of 2D defect centers may be improved by integrating the host material with conventional semiconductor materials\cite{quellmalz2021large} or via van der Waals encapsulation\cite{liu2022scalable} -- a technique which can also be used to tune the properties of 2D defects\cite{su2022tuning}. Lastly, the spin coherence time -- a key parameter for quantum technology applications -- can be significantly larger for defects in 2D materials, due to the lower density of surrounding nuclear spins\cite{onizhuk2021substrate,sajid2022spin, ye2019spin}. 

So far, relatively few 2D defect systems have been explored in experiments. The most widely studied 2D host materials include hBN \cite{tran2016quantum,sajid2020single, sajid2020theoretical, ivady2020ab, patel2022probing}, SiC \cite{bekaroglu2010first, hashemi2021photoluminescence}, and the semiconducting transition metal dichalcogenides (TMDs)\cite{bertolazzi2017engineering}. With the important exception of C and O-related defects in hBN, the study of point emitters in these materials have mainly focused on intrinsic defects. As a consequence, it remains of great interest to investigate other 2D defect systems, and establish the extent to which they may exhibit superior properties. Such investigations represent a daunting challenge for experiments, which are often complex, time-consuming, and indirect (with regards to the atomic nature of the emitter). On the other hand, first-principles calculations can be used to screen thousands of defect structures to identify the most promising candidates before they are explored in the lab\cite{bertoldo2022quantum,davidsson2023absorption}. 

Point defects in insulating crystals may form localised quantum states within the band gap, which upon excitation can result in the emission of photons. The electronic transitions at the defect center couple to local lattice vibrations. This coupling will be imprinted on the line shape of the PL spectrum producing a unique fingerprint -- not only for the defect structure, but also for the specific electronic transition. Consequently, calculation of the PL line shape from first principles is a powerful means to characterise point defects. Moreover, a symmetry analysis of localized defect states can be used to derive optical selection rules and the polarization distribution of the emitted photons. Because most of the potential applications of point defects in quantum information technology utilise the spin degree of freedom, defects with high-spin ground states are of particular interest.\cite{wolfowicz2021quantum} In this context, not only the optical properties, but also the hyperfine (HF) and zero-field splitting (ZFS) tensors, $g$-factors, spin-orbit coupling, and spin coherence times, are of key importance. Such spin-related quantities can also be obtained with good precision from first-principles calculations.\cite{ivady2018first,bertoldo2022quantum,sajid2020single}     

In this work we employ a workflow of first-principles calculations to screen for 2D point defect systems with triplet ground states, long spin coherence times, and narrow emission line shapes. The systems explored comprise ten non-magnetic, large-band gap 2D host materials with low density of nuclear spin isotopes. Within these host materials we create a total of 5388 intrinsic and extrinsic point defects with an even number of electrons in either the neutral or $\pm 1$ charge states. Our density function theory (DFT) calculations show that 596 of these defects, i.e. about 10\%, feature a triplet ground state. After screening the electronic excitations (in both spin channels) of these defects for the excitation energy and electron-phonon coupling strength, we identify 39 unique transitions with zero phonon line (ZPL) energy between 0.2 and 4.0 eV and Huang-Rhys (HR) factors below 10. Some of the most promising defects are discussed in more detail. All the results will be available as part of the open QPOD database\cite{bertoldo2022quantum}. 

\section{Results and discussion}\label{sec:results}
\begin{figure*}
  \centering
  \includegraphics[width=1.\linewidth]{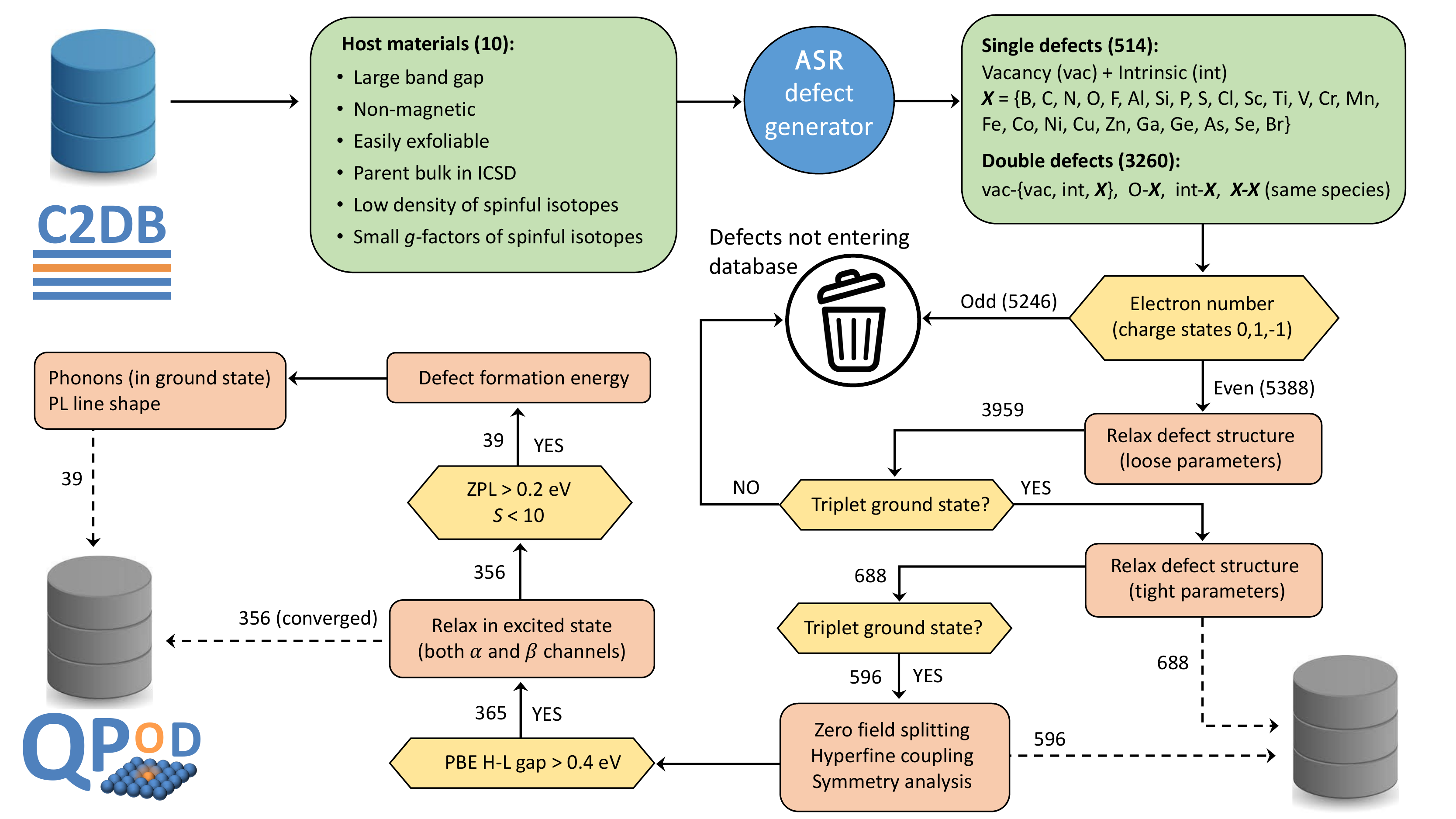}
  \caption{\textbf{Workflow for high-throughput search of triplet defects.} Starting from the C2DB database, 10 monolayers materials are selected as defect hosts based on properties of the pristine material such as the size of band gap, non-magnetic ground state, low abundance of spinful isotopes, thermodynamic stability, and existence of a known layered bulk parent. Subsequently, the ASR defect builder is used to generate all inequivalent point defects comprising vacancies, substitutions (intrinsic and extrinsic), vacancy + substitutions, double vacancies, and double substitutions. Defect systems with an even number of electrons (candidates for triplets) are selected for DFT calculations with loose computational parameters to identify the systems with triplet ground state. For the predicted triplets, calculations with tight parameters are then performed. For the 610 confirmed triplets, the zero field splitting, hyperfine coupling, and orbital symmetries are calculated. Excited state calculations in both spin channels are then performed for the triplets with PBE HOMO-LUMO gap larger than 0.4 eV. For the defects with Huang-Rhys (HR) factors below 10.0, the full photoluminescence (PL) spectrum is calculated. All results generated with tight computational parameters will be uploaded to the QPOD database \cite{bertoldo2022quantum}.}

  \label{fig:1.1_workflow}
\end{figure*}

A graphical overview of the workflow is shown in Figure \ref{fig:1.1_workflow}. All density functional theory (DFT) calculations were performed using the GPAW electronic structure code\cite{enkovaara2010electronic}. The workflow was developed using the Atomic Simulation Recipes (ASR)\cite{gjerding2021atomic} -- a Python framework for constructing and orchestrating computational workflows with automatic logging of metadata and task dependencies to ensure reproducibility of the results. The tasks of the ASR workflow were submitted using MyQueue\cite{mortensen2020myqueue}, which is a front end to common job schedulers supporting SLURM, LSF, and PBS. 

\subsection{Host materials}
As a first step we select the 2D crystals to be considered as host materials for the point defects. For a defect state to be addressable by electromagnetic fields it needs to be spatially localized and energetically isolated from the bulk bands of the host system. This calls for host materials with sufficiently large band gaps. To meet this condition we require host materials to have a band gap of at least 2 eV when calculated with the HSE06 xc-functional, which is known to reproduce band gaps of 2D materials reasonably well\cite{haastrup2018computational}. In order to maximise the likelihood that the proposed defect systems can be realised, we furthermore restrict ourselves to 2D hosts which have already been synthesized, at least in their bulk form, or whose energy lies within 50 meV/atom of the convex hull. The latter condition means that the considered structure has an internal energy of at most 50 meV/atom compared to the most stable known phase of its elements. Among all the materials considered only GeS does not appear in one of the experimental crystal structure databases ICSD or COD. Since long coherence times is beneficial in many applications of spin defects\cite{wolfowicz2013atomic, miao2020universal}, we only consider host systems with low natural concentration of spinful isotopes and/or small $g$-factors. We note that recent work has demonstrated that the coherence time of spin defects are largely determined by the properties of the nuclear spin bath in the host material\cite{kanai2022generalized,sajid2022spin}. Finally, we exclude materials with heavy elements as these have strong spin-orbit coupling rendering the defect spin less well defined. 
After screening the Computational 2D Materials Database (C2DB)\cite{haastrup2018computational, gjerding2021recent} against these criteria, we end up with the ten host materials listed in Table \ref{table:hosts}.

\begin{table*}[h]
    \centering
    \begin{tabular}{cccccccc}
    \hline \hline         \\
&Host crystal &C2DB UID &Layer group &$\Delta H_{\mathrm{{hull}}}$ [eV/atom]  & $E_{\mathrm{gap}}^{\mathrm{HSE06}}$ [eV] &ICSD/COD ID      \\
\hline
&BN    &BN-4a5edc763604  &$p-6m_{2}$     &0.000    &5.68  &186248                                 \\
&MoS$_2$  &MoS2-b3b4685fb6e1    &$p-6m_{2}$    &0.000  &2.09   &38401                            \\
&MgI$_2$  &MgI2-440644551de5     &$p-3m_{1}$    &0.001  &4.20  &281551                     \\
&MgBr$_2$  &MgBr2-702277c8c7ed    &$p-3m_{1}$     &0.000  &5.74 &9009017                             \\ 
&GeS &Ge2S2-ecbb7c185669 &$pm2_1n$    &0.031  &2.37  & N/A                   \\
&SiS$_2$   &Si4S8-69fbe379c2cf      &$p2_{1}/b11$   &0.010  &4.55  &291212                                 \\
&CaI$_2$    &CaI2-ee886d522d75      &$p-3m_{1}$   &0.000  &4.81   &9009097                           \\
&V$_{2}$O$_{5}$   &V4O10-3095e0377954 &pmmm  &0.000    &4.47 &4124511                                  \\
&ZnCl$_2$       &ZnCl2-1b7175e04416   &$p-4m_{2}$  &0.000  &5.70 &8103829                            \\
&YCl$_{3}$   &Y2Cl6-7a926d84d193      &$p-31m$   &0.000   &6.73 &1528118 \\
\hline  
    \end{tabular}

    \caption{\textbf{2D Host materials.} The table shows the chemical formula of the material, the unique identifier in the C2DB database, the layer group (2D analogue of the space group), the energy above the convex hull, the band gap calculated with the HSE06 exchange-correlation functional, and the identifier of the parent bulk crystal in the ICSD or COD crystal structure databases.}
    \label{table:hosts}
\end{table*}

\subsection{Generation of point defects}
Point defects in the ten host materials are generated using the ASR defect builder\cite{davidsson2023absorption}. We include the following single-site defects: vacancies ('vac'), intrinsic ('int') substitutions, and extrinsic ('X') substitutions. The following extrinsic dopants are included: X$\in $\{B, C, N, O, F, Al, Si, P, S, Cl, Sc, Ti, V, Cr, Mn, Fe, Co, Ni, Cu, Zn, Ga, Ge, As, Se, Br\}. The X-dopants comprise most elements from period 2-4. The selected elements are relatively light and have small atomic radius. These criteria were chosen in order to minimise effects of spin-orbit interactions and for size compatibility with the host materials. 

In addition, we include the following two-site defects by combining single-site defects on neighboring atomic sites: vac+vac, vac+int, vac+X, oxygen+X, int+X, X+X (only the same external element on the two neighboring sites). We did not consider interstitial defects. 

Each of the generated defects are considered the neutral and $\pm 1$ charge states. This yields a total of 10.634 unique defects across the ten host materials. The defects with an even number of electrons (candidates for triplet ground state) are relaxed using the PBE xc-functional with "loose" parameter settings, i.e. the $\Gamma$-point only and 500 eV plane wave cut off. The defects are represented in a supercell that breaks the symmetry of the host crystal and ensures a defect-defect separation of at least 15 \AA, see Ref. \cite{bertoldo2022quantum}. The loose DFT parameters ensure a computationally efficient means of identifying the systems with a triplet ground state. This approach is meaningful because (i) a triplet ground state is a rare event, and (ii) the defect spin state is reliably reproduced with the loose parameter settings. Finally, DFT relaxations with tight parameters, i.e. $k$-point density of 3 $\textrm{\AA}$ and 800 eV plane wave cut-off, are performed for the defects predicted to have a triplet ground state. Out of 688 triplets predicted with the loose settings, 596 are confirmed using the tight parameters.

\begin{table}[h]
    \centering
    \begin{tabular}{cccccccccccc}            
\hline \hline
&Material & $T_2 (\mathrm{ms})$ & $N_\mathrm{even}$  & $\%$ of Triplets \\
\hline
&BN      & 0.045 &398  & 7.29                                 \\
&MoS$_2$  & 2.50 &559  & 18.60                            \\
&MgI$_2$  & 0.11 &557  & 22.80                              \\
&MgBr$_2$  & 0.14 &494  & 20.45                                \\ 
&GeS & 10.0 &563  & 10.66                          \\
&SiS$_2$   & 40.0 &384  & 10.94                                 \\
&CaI$_2$  & 0.14 &352  & 18.75                            \\
&V$_{2}$O$_{5}$ & 0.046 &324  & 2.47                                   \\
&ZnCl$_2$ & 0.77 &262  & 15.65                               \\
&YCl$_{3}$ & 0.76 &66 & 27.27  \\
  \hline \hline
    \end{tabular}
    \caption{\textbf{Defect spin statistics.} The spin coherence time
    $T_2$, total number of defects with even electron number (numbers representing the  relaxed defects  with loose parameters) as well as the percentage of these that are triplets for the ten host materials. The low percentage of triplets in V$_2$O$_5$ is related to the poor convergence rate of the relaxations with tight parameters for this host material.}
    \label{table:triplet_stats}
\end{table}

\subsection{Spin properties}
 The spin coherence time, $T_2$, of a defect spin is governed by the hyper fine (HF) coupling tensor and the dipole-dipole interactions between the nuclear spins with strength given by the Landé $g$-factors. From these parameters, $T_2$ can be calculated using the cluster-correlation expansion method\cite{onizhuk2021pycce}. It has been found that the $T_2$ is insensitive to the atomistic details of the defect center and rather is dictated by the nuclear-spin properties of the host material\cite{sajid2022spin,ye2019spin}. We therefore limit ourselves to calculation of spin coherence times for a representative defect in each host material. 

\begin{figure}
    \centering
    \includegraphics[width=1.\linewidth]{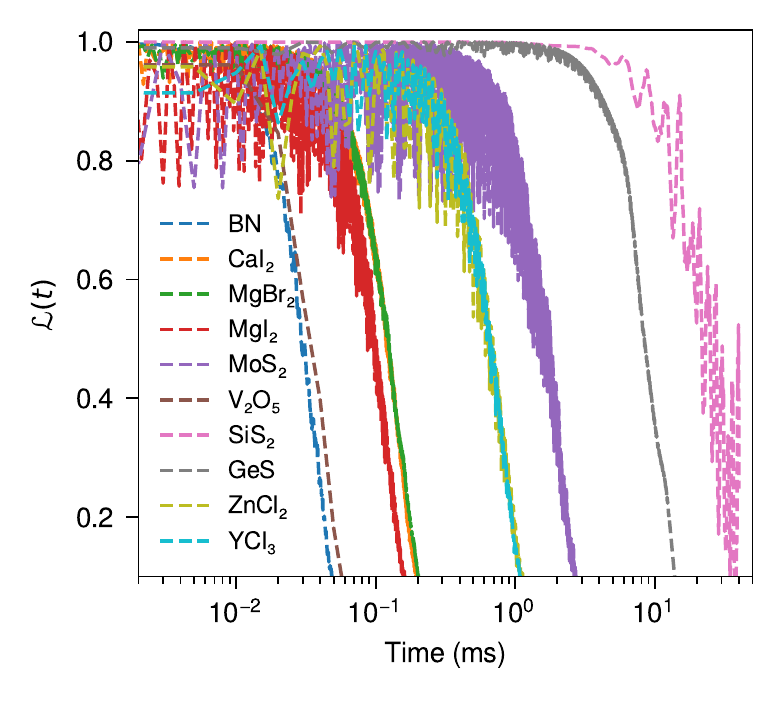}
    \caption{\textbf{Spin coherence times}. The calculated coherence function, $\mathcal{L}(t)$, for a representative defect within each of the ten host materials. The spin coherence time is obtained by fitting the exponential decay of $\mathcal{L}$ in the long time limit, see Table \ref{table:triplet_stats}.}
    \label{fig:Coherence}
\end{figure}

 In Figure \ref{fig:Coherence} we plot the calculated coherence function $\mathcal{L}(t)$ for a hypothetical defect (cation vacancies in $S=1$ spin state, since results are not sensitive to the geometrical structure of the defect) in all the host materials. The coherence function is fitted by the exponential function exp$(-t/T_2)^n$\cite{sajid2022spin} to extract the spin coherence time, $T_2$. In Table \ref{table:triplet_stats} we list the spin coherence times, together with the number of defects with even electron number and the percentage of these which are triplets for all host materials. We note that the defects in three of the host materials (MoS$_2$, GeS, and SiS$_2$) have particularly long spin coherence times (2.5ms, 10ms and 40ms, respectively) well above that of the N$_{V}^{-1}$ centre in diamond (1.0ms).  These long spin coherence times stem from the low concentration of spinful nuclear isotopes in the host materials. 
 
 The zero field splitting (ZFS) is calculated from the Kohn-Sham eigenstates while the HF coupling tensor is calculated from the electron spin-density (for details see Ref. \cite{bertoldo2022quantum}). The calculated value of the ZFS parameters ($D$ and $E$ components) are listed in Table \ref{table:PL_results}. Application of the defects in an ODMR setup requires optical detection of the spin resonance, which is produced through  microwaves with frequency matching the splitting between magnetic sub-levels. Therefore, the ZFS should be within the experimentally achievable frequency range. We note that the magnitude of $D$ (D the axial component of the magnetic dipole–dipole interaction) for all the defects (with the exception of  $\mathrm{Cr_N}^{-1}$ in hBN and $\mathrm{v_ICo_{Mg}}$ in MgI$_2$) is below 10GHz, which is well within the experimentally accessible range \cite{PILBROW19961465}. We also note that our calculated ZFS values for the $\mathrm{v}_\mathrm{B}^-$ defect in hBN of $D=2.97$Gz and $E=0.0$Gz are in good agreement with the experimentally measured values\cite{gottscholl2020initialization} of 3.5GHz and $0.005$GHz, respectively.

\subsection{Excited states}
\begin{figure}[t]
    \centering
    \includegraphics[width=.65\linewidth]{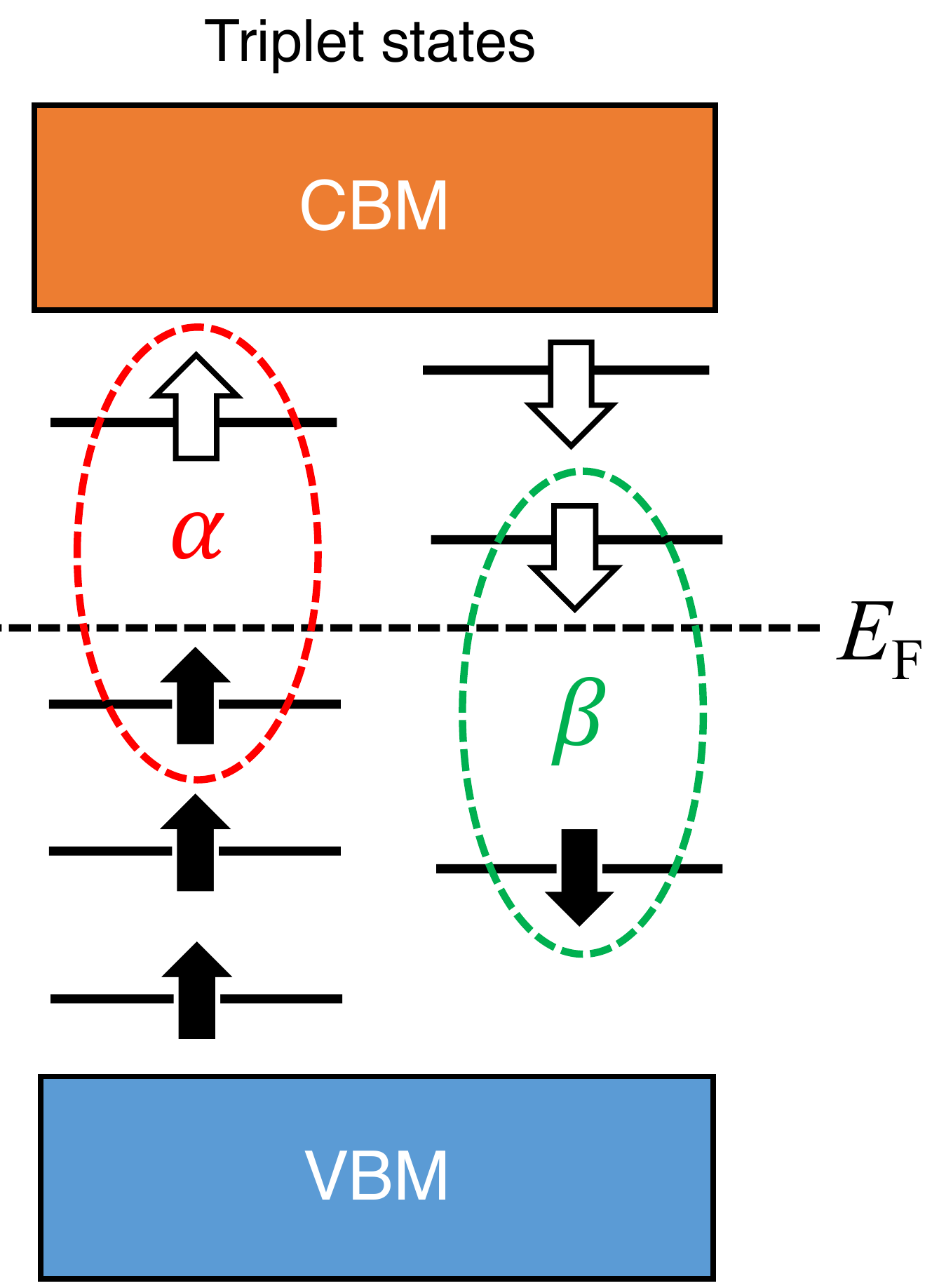}
    \caption{\textbf{HOMO-LUMO excitations in a triplet.} Typical spin configuration of a high-spin (triplet) defect system. The HOMO-LUMO excitations within the $\alpha$ and $\beta$ spin channels are marked by red and green circles, respectively}. 
    \label{Triplet_humo_lumo_excitations}
\end{figure}

For each defect with a triplet ground state and an energy gap between the highest occupied molecular orbital (HOMO) and lowest unoccupied molecular orbital (LUMO) of at least 0.4 eV (PBE Kohn-Sham gap), we perform DFT calculations with non-aufbau occupations corresponding to the lowest transitions in both the majority ($\alpha$) and minority ($\beta$) spin channels, see Figure \ref{Triplet_humo_lumo_excitations}. These excited state calculations are performed using the direct-optimization of maximally overlap orbitals (DO-MOM)\cite{ivanov2021method, levi2020variational} method as implemented in the GPAW code. This method is based on direct optimization (DO) of orbital rotations by means of an efficient quasi-Newton method for saddle points, in combination with the maximum overlap method (MOM). The MOM ensures that the character of the states is preserved during the optimization procedure. The DO-MOM has previously been shown to give better convergence behavior for defect states than the traditional Delta-SCF method\cite{bertoldo2022quantum}.

\begin{figure}[t]
    \centering
    \includegraphics[width=1.\linewidth]{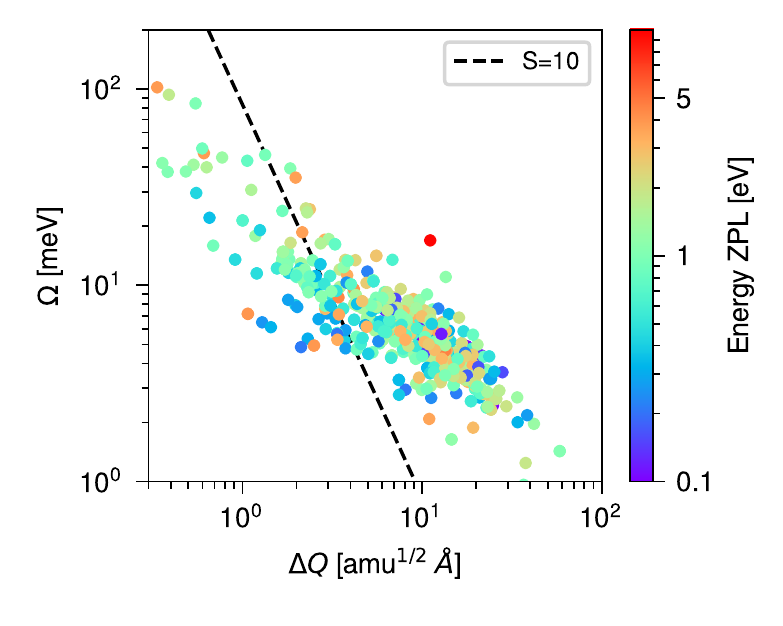}
    \caption{\textbf{Trends in vibronic coupling.} The average frequency, $\Omega$, estimated from the curvature of 1D configuration coordinate diagrams versus the mass weighted difference in the geometrical coordinates of the ground and excited state, $\Delta Q$. Defects below the dashed black line have an effective Huang-Rhys factor, $S$, below 10. The ZPL energy of the transition is indicated by the color code}. 
    \label{HR_factors_1D_CCD}
\end{figure}

After creating the non-aufbau occupations for each spin channel, the structures are relaxed using the ASR workflow for excited states employing the DO-MOM method. The zero-phonon line (ZPL) energies are obtained as the difference between the DFT total energy in the ground state and the relaxed excited state. In addition, we calculate the mass weighted difference in the geometrical coordinates of ground and excited states, 
\begin{equation}
\Delta Q^2 = \sum_\alpha M_\alpha \Delta R_\alpha^2,
\end{equation}
where $M_\alpha$ is the mass and $\Delta R_\alpha$ the displacement of atom $\alpha$, and the sum runs over all the atoms in the defect supercell. 

\begin{figure}[t]
    \centering
    \includegraphics[width=1.\linewidth]{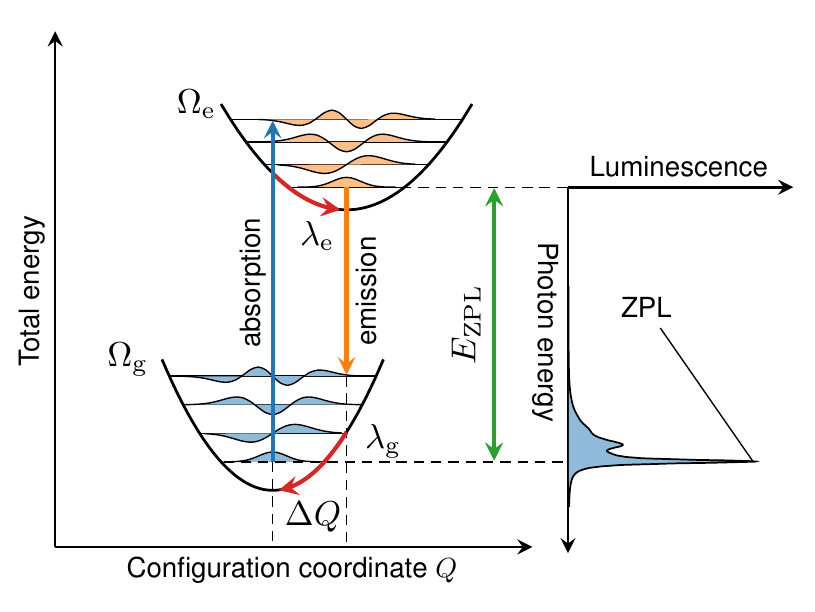}
    \caption{\textbf{Configuration coordinate diagram.} Left: One-dimensional configuration coordinate diagram with an absorption and emission process indicated by blue and orange arrows. 
    The zero phonon line energy ($E_{\mathrm{ZPL}}$), reorganisation energies in the ground and excited state ($\lambda_\mathrm{g}$ and $\lambda_{\mathrm{e}}$), and the mass weighted displacement ($\Delta Q$) are illustrated. Right: A photoluminescence spectrum with a ZPL and phonon side bands.} 
    \label{fig:ccdiamgrame}
\end{figure}

Next, we use the one-dimensional configuration coordinate diagram (1D-CCD) method to estimate the Huang-Rhys (HR) factor. The HR factor quantifies the number
of phonons emitted during the transition. In the 1D-CCD, an effective vibrational frequency, $\Omega$, is defined from the curvature of the 1D parabola in the electronic ground state, see Figure \ref{fig:ccdiamgrame}. The HR factor is then obtained as 
\begin{equation}
S = \frac{\lambda_{g}}{\hbar \Omega}  
\end{equation}
where $\lambda_{g}$ is the reorganisation energy in the ground state. Figure \ref{HR_factors_1D_CCD} shows $\Omega$ plotted against $\Delta Q$ for all the triplet transitions. The ZPL is indicated by the color code. The dots below the dashed black line represents transitions with $S < 10$.

\subsection{Defect formation energies}
The stability of the defects is assessed by their formation energy, $E^{\mathrm{f}}$. For neutral defects, the formation energy is first calculated with respect to the chemical potential of the elements in their standard states $\mu^\mathrm{ref}$, extracted from the OQMD database \cite{saal2013materials, kirklin2015open}. 
For the binary compounds ($A_x B_y$) considered in the present study, B-poor conditions are defined as $\mu_A=\mu_A^{\mathrm{ref}}$ and $\mu_B = \Delta H(A_xB_y)/y + \mu_B^{\mathrm{ref}}$, where $\Delta H$ is the heat of formation of the host material\cite{haastrup2018computational}. A similar expression holds for A-poor conditions. We evaluate $E^{\mathrm{f}}$ under both A-poor and B-poor conditions using the reference chemical potential (standard state) for extrinsic elements.

The formation energy of charged defects is obtained by correcting the neutral formation energy by the charge transition energy. We calculate the latter using Slater-Janak (SJ) theory as outlined in Ref. \cite{bertoldo2022quantum}. The main advantage of the SJ approach compared to the standard total energy approach is that it avoids comparing total energies of systems with different charges and thus avoids the need for electrostatic correction schemes\cite{freysoldt2011electrostatic}. It should be noted that for charged defects the formation energy depends on the position of the Fermi energy. The latter is in general unknown, but can in principle be calculated self-consistently in conjunction with the concentration of relevant (charged) defects, which are usually assumed to follow a Boltzmann distribution\cite{bertoldo2022quantum}. In this work, for simplicity and because we do not assume an equilibrium defect distribution, we do not attempt to determine the Fermi level, but simply set it to the centre of the band gap.

The formation energies for the final set of defects listed in Table \ref{table:PL_results} ranges from -1.76 to 13.3 eV. Defects with lower formation energy are expected to be more stable, which is an important requirement for applications. We note that the $\mathrm{v}_\mathrm{B}^-$ defect in hBN has a formation energy around 8 eV (depending on the B and N chemical potentials) and shows good stability in experiments at room temperature. This indicates that even defects with relatively large formation energies can be experimentally realised, e.g. by irradiation or implantation. For this reason we do not discard defects with high formation energy. The $\mathrm{O_SV_{Si}^{-1}}$ defect in $\mathrm{SiS_2}$ has a negative formation energy, which signals that the host material is unstable with respect to formation of this defect. However, the formation energy was calculated with respect to the standard states of the extrinsic elements, corresponding to high concentrations. Depending on the experimental technique used to engineer the defects the concentration of extrinsic elements can be kept low, for which case the formation energy would remain positive.

\subsection{Photo-luminescence line shape}
\begin{figure}[t]
    \centering
    \includegraphics[width=0.9\linewidth]{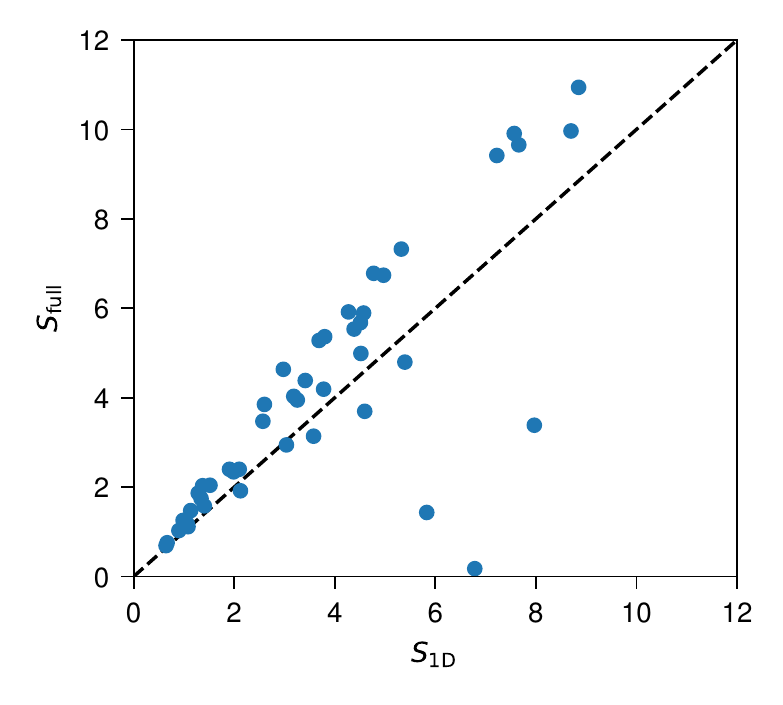}
    \caption{\textbf{Huang-Rhys factors: 1D model vs. full phonon method},  A comparison of HR factors using the 1D-configuration coordinate diagram method versus the full phonon method, for all the HOMO-LUMO transitions in the $\alpha$ and $\beta$ spin channels. There is a reasonable agreement between the two methods for majority of the cases}. 
    \label{fig:2_2.2_formation}
\end{figure}

For the triplet defects with zero phonon line (ZPL) larger than 0.2 eV and $S<10$, we calculate the photo luminescence (PL) spectrum following Ref. \cite{alkauskas2014first}. The key ingredient of the calculations are the $\Gamma$-point phonons of the supercell in the electronic ground state, which allow us to obtain the partial HR factors  
\begin{equation}
S_k  = \frac{1}{2\hbar} \omega_k \Delta Q_k^2, 
\label{Eq:Sk}
\end{equation}
where $\Delta Q_k$ are the coordinates of the displacement vector in the normal modes, i.e. $\Delta Q = \sum_k \Delta Q_k$. The electron-phonon spectral function is defined as
\begin{equation}
S(\omega)  = \sum_k S_k \delta(\omega - \hbar \omega_k), 
\end{equation}
which can be integrated to yield the total HR factor of the transition. Although this will in general differ from the HR factor of the 1D-CCD there is a reasonable agreement between the two methods (see figure \ref{fig:2_2.2_formation}). Finally, the PL spectrum is evaluated by applying a generating function to the electron-phonon spectral function\cite{alkauskas2014first}.

\begin{figure*}
    \centering
    \includegraphics[width=0.9\linewidth]{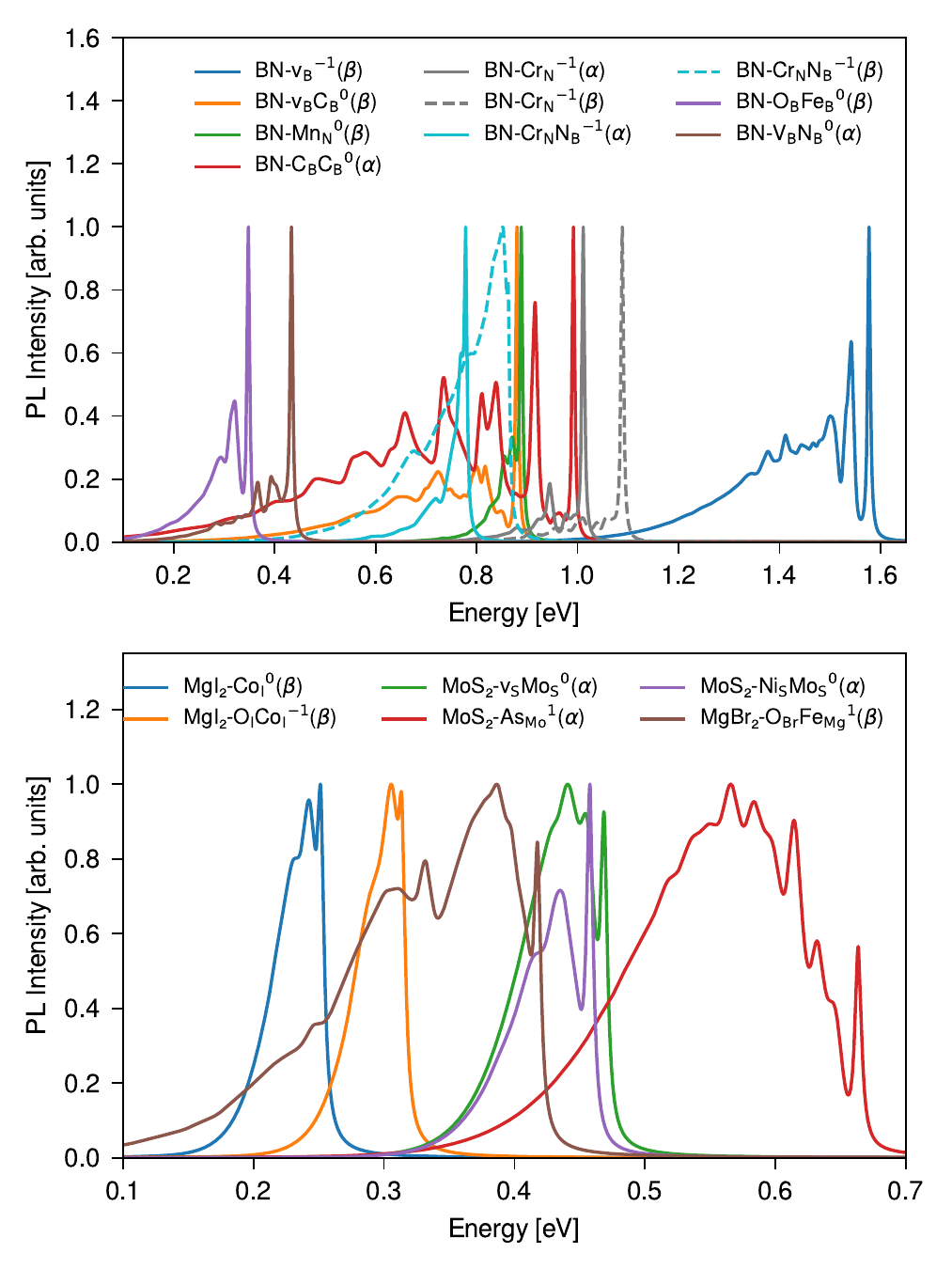}
    \caption{\textbf{PL lineshapes.} Calculated PL lineshapes of selected triplet defects with ZPL $<$ 2.0 eV and HR factor $<$ 3.5. Guassian smearing with a width of 3meV is used to broaden the calculated spectra. The HOMO-LUMO transitions in majority ($\alpha$) and minority ($\beta$) spin channel are plotted. The charge states are mentioned on the defect notation.  The Intensities are normalised with respect to the maximum intensity of the line shapes. We find new sharp single photon emitting defect centres in hBN, MgI$_{2}$, MoS$_2$.}
    \label{fig:PL_lineshape}
\end{figure*}

Figure \ref{fig:PL_lineshape} shows the PL line shapes of selected defects with ZPL below 2.0 eV and HR factor $S<3.5$. Table \ref{table:PL_results} lists the details of all 40 transitions resulting from the screening including the ZPL energy, $\Delta Q$, $S$, the point group of the defect systems, the symmetry of the HOMO and LUMO levels, the zero-field splitting, and the formation energy of the defect. The HR threshold of 3.5 used to select the defects in Figure \ref{fig:PL_lineshape} corresponds to that of the well known single photon emitter NV$^{-1}$ in diamond\cite{alkauskas2014first}. 

\begin{figure}[t]
    \centering
    \includegraphics[width=0.9\linewidth]{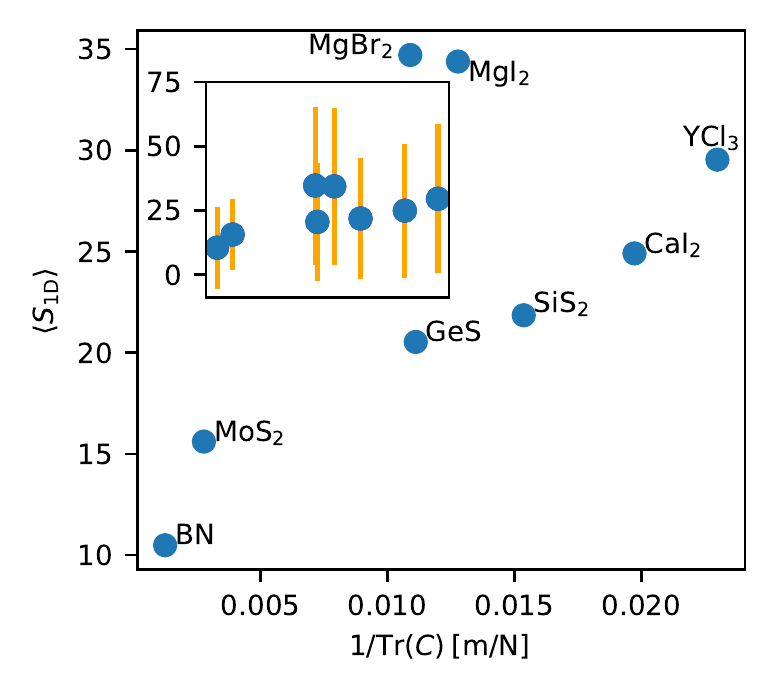}
    \caption{\textbf{Huang-Rhys factor as function of stiffness.} Average HR factor $S$ as a function of the inverse of the trace of the stiffness tensor. Due to the poor convergence of the excited states for V$_2$O$_5$ and ZnCl$_2$ these materials were excluded from the figure. The inset shows the same data but also includes the standard deviation of $S$. The HR factors were computed using the 1D-Configuration coordinate diagram method. We only averaged over defects with $S<100$.}. 
    \label{fig:Svsstiffness}
\end{figure}

One can see from Table \ref{table:PL_results} and Figure \ref{fig:PL_lineshape} that hBN is the most frequently occurring host material followed by MoS$_2$. It follows from Table \ref{table:triplet_stats} that the reason is not a higher propensity for the defects in these materials to form triplet ground states. Instead, the reason is a generally lower HR factor found for the defects in hBN and MoS$_2$. We ascribe this to the high stiffness (in particular for hBN) compared to the other materials in this study. A material with higher stiffness  is expected to have smaller $\Delta Q$ and consequently also a smaller HR factor (see Eq. \ref{Eq:Sk}). 
The importance of the stiffness is further illustrated Figure \ref{fig:Svsstiffness} where we show the average HR-factor as a function of the inverse of the trace of the stiffness tensor. It is clear that materials with higher stiffness have a higher probability to host defects with small HR factors. 
This relation between the stiffness of the host material and the HR factor of its defects, therefore constitutes an important guideline when searching for novel defect systems with bright and narrow emission spectra. 
However, as can be seen in the inset of the same figure, the variance  of the HR factors is large, which indicates that the local properties of the defects also has a large influence on the HR factor. This variance might also partly explain why the Mg compounds deviate from the trend and have high HR factors compared to their relatively high stiffness.

Our calculated ZPL energies, PL line shapes and HR factors for the V$_{\mathrm{B}}^{-1}$ defect are in good agreement with previous studies\cite{sajid2020vncb, SajidAli2020Edgeeffects, linderalv2021vibrational}. We find new sharp triplet emitters, e.g. the Cr$_{\mathrm{N}}^{-1}$ defect in hBN for transitions in both the $\alpha$ and $\beta$ spin channels with HR factors of 0.76 and 0.69, respectively. The analysis of spectral density reveals that the dominant contribution to the HR factor of the Cr$_{\mathrm{N}}^{-1}$ defect comes from coupling to a single phonon mode at 66 meV (for the $\alpha$ transition) and to two phonon modes at 27 meV and 84 meV (for $\beta$ transition), respectively. As can be noted from Table \ref{table:PL_results}, the Cr$_{\mathrm{N}}^{-1}$ defect with C$_{3\mathrm{V}}$ point group would emit photons with polarisation oriented perpendicular and parallel to the hBN plane for the $\alpha$-transition in (e $\rightarrow$ a$_{1}$) and $\beta$-transition (a$_{1}$ $\rightarrow$ a$_{1}$), respectively. For MoS$_2$, only the $\mathrm{v_SMo_S}^{0}$  and $\mathrm{Ni_SMo_{S}}^{0}$ defects show sharp transitions (in the $\alpha$-channel) with HR factor of 2.34 and 1.74, respectively. These HR factor arise mainly due to coupling with low energy phonon modes at 13 meV and 18 meV, respectively. Finally, MgBr$_{2}$ and CaI$_2$ host one sharp emission center each.

\begin{table*}[h]
    \centering
    \begin{tabular}{ccccccccc}
    \hline \hline 
    \setlength{\dashlinedash}{0.2pt}
    \setlength{\dashlinegap}{4.5pt}
    \setlength{\arrayrulewidth}{0.2pt}

Host &Defect(spin) &$\Delta Q$  &ZPL   &$S$    &Point group (H/L) &ZFS (\textit{D/E}) &$E^\mathrm{f}(\mathrm{A-poor})$  &$E^\mathrm{f}(\mathrm{B-poor})$\\
\hline
BN &$\mathrm{v_B}^{-1}(\alpha)$  &0.61  &3.79 &2.40   &$\mathrm{D_{3h}}$($\mathrm{e^{\prime}}_{1}$/$\mathrm{e^{\prime\prime}}_{1}$)    &2.97/0.0 &7.35 &8.64 \\
&$\mathrm{v_B}^{-1}(\beta)$    &0.63   &1.58   &2.40  &$\mathrm{D_{3h}}$($\mathrm{e^{\prime\prime}}_{1}$/$\mathrm{e^{\prime\prime}}_{1}$) &$2.97/0.0$ &$7.35$ &$8.64$\\		
&$\mathrm{v_BC_B}^{0}(\alpha)$ &0.48  &3.81  &2.03    &$\mathrm{C_{1}}$($\mathrm{a}$/$\mathrm{a}$) &0.54/0.03 &8.01 &10.59\\ 
&$\mathrm{v_BC_B}^{0}(\beta)$  &0.47  &0.88  &1.92    &$\mathrm{C_{1}}$($\mathrm{a}$/$\mathrm{a}$) &0.54/0.03 &8.01 &10.59\\ 

&$\mathrm{Mn_N}^{0}(\alpha)$ &1.12 &1.52 &3.70    &$\mathrm{C_{3V}}$($\mathrm{a_{1}}$/$\mathrm{e}$)   &-4.14/-1.26 &8.42 &7.13 \\
&$\mathrm{Mn_N}^{0}(\beta)$   &0.69 &0.89  &1.03    &$\mathrm{C_{3V}}$($\mathrm{e}$/$\mathrm{a_{1}}$)  &-4.14/-1.26  &8.42 &7.13  \\

&$\mathrm{Cr_N}^{-1}(\alpha)$  &0.38   &1.01  &0.76        &$\mathrm{C_{3V}}$($\mathrm{e}$/$\mathrm{a_{1}})$   &25.69/0.06  &10.57 &9.27 \\
&$\mathrm{Cr_N}^{-1}(\beta)$  &0.36  &1.09    &0.69    &$\mathrm{C_{3V}}$($\mathrm{a_{1}}$/$\mathrm{a_{1}})$   &25.69/0.06   &10.57 &9.27\\
&$\mathrm{Cr_NN_B}^{-1}(\alpha)$  &1.06  &0.78 &1.43     &$\mathrm{C_{1}}$($\mathrm{a}$/$\mathrm{a})$   &-6.78/-0.66   &11.81 &13.10\\
&$\mathrm{Cr_NN_B}^{-1}(\beta)$  &1.67  &0.86   &3.38   &$\mathrm{C_{1}}$($\mathrm{a}$/$\mathrm{a}$)    &-6.78/-0.66  &11.81  &13.10\\
&$\mathrm{Cr_NN_B}^{1}(\beta)$   &1.18   &1.31 &4.63   &$\mathrm{C_{1}}$($\mathrm{a}$/$\mathrm{a}$) &3.0/0.95 &11.96 &13.25\\
&$\mathrm{O_BFe_B}^{0}(\beta)$   &0.66 &0.35   &1.48     &$\mathrm{C_{1}}$($\mathrm{a}$/$\mathrm{a}$) &-1.31/-0.40 &9.0 &11.59\\
&$\mathrm{Mn_BB_N}^{0}(\alpha)$   &1.82  &0.93   &4.80    &$\mathrm{C_{1}}$($\mathrm{a}$/$\mathrm{a}$) &3.36/1.03 &9.47 &8.18 \\
&$\mathrm{V_BN_B}^{0}(\alpha)$   &0.55  &0.43   &1.11    &$\mathrm{C_{1}}$($\mathrm{a}$/$\mathrm{a}$) &8.17/1.72 &8.11 &11.99\\ 
\hline
\\[-1em]
MgI$_2$  &$\mathrm{Co_{Mg}}^{-1}(\beta)$  &1.68  &1.89  &5.89   &$\mathrm{D_{3d}}$($\mathrm{e_{g}}$/$\mathrm{a_{1g}}$) &2.20/0.58 &2.11 &3.19\\
&$\mathrm{Co_I}^{0}(\beta)$    &1.32  &0.25  &1.87   &$\mathrm{C_{3V}}$($\mathrm{a_{1}}$/$\mathrm{e}$) &-0.22/-0.04 &5.26 &4.72\\
&$\mathrm{v_ICo_{Mg}}^{0}(\beta)$  &2.31  &0.30 &4.38  &$\mathrm{C_{1}}$($\mathrm{a}$/$\mathrm{a}$) &17.54/5.59 &3.71 &4.25\\
&$\mathrm{v_ICo_I}^{1}(\beta)$   &2.13  &0.21 &3.85  &$\mathrm{C_{1}}$($\mathrm{a}$/$\mathrm{a}$) &-2.43/-0.61 &8.67 &7.59\\
&$\mathrm{O_ICo_I}^{-1}(\beta)$   &1.55 &0.31  &2.04    &$\mathrm{C_{1}}$($\mathrm{a}$/$\mathrm{a}$) &1.75/0.11 &4.38 &3.30\\
\hline
\\[-1em]
MoS$_2$  &$\mathrm{v_SMo_S}^{0}(\alpha)$   &1.21  &0.47  &2.34 &$\mathrm{C_{3V}}$($\mathrm{e}$/$\mathrm{a_{1}}$) &8.57/0.01  &8.70 &6.86\\
&$\mathrm{Co_{Mo}}^{1}(\beta)$    &1.90  &0.74   &6.74 &$\mathrm{C_{1}}$($\mathrm{a}$/$\mathrm{a}$) &0.98/0.32   &3.40 &4.32  \\
&$\mathrm{v_{S}Al_{Mo}}^{1}(\alpha)$   &1.81  &0.42   &4.99 &$\mathrm{C_{s}}$($\mathrm{a}$/$\mathrm{a}$) &0.33/0.01 &4.32 &4.78 \\
&$\mathrm{v_{S}C_{Mo}}^{0}(\alpha)$   &2.55 &0.33 &9.97 &$\mathrm{C_{s}}$($\mathrm{a}$/$\mathrm{a}$)   &0.34/0.02  &8.01  &8.47  \\
&$\mathrm{v_{S}C_{Mo}}^{0}(\beta)$   &1.71  &0.44   &5.53 &$\mathrm{C_{s}}$($\mathrm{a}$/$\mathrm{a}$) &0.34/0.02  &8.01 &8.47\\
&$\mathrm{As_{Mo}}^{1}(\alpha)$   &1.01   &0.66   &3.47    &$\mathrm{D_{3h}}$($\mathrm{e^{\prime}}_{1}$/$\mathrm{a^{\prime}}_{1})$        &0.64/0.01   &5.28 &6.21  \\
&$\mathrm{O_SAl_{Mo}}^{-1}(\alpha)$  &1.60 &1.14  &5.37  &$\mathrm{C_{s}}$($\mathrm{a}$/$\mathrm{a}$)  &0.48/0.05 &0.44 &0.90            \\
&$\mathrm{O_SCu_{S}}^{-1}(\alpha)$    &1.99  &0.68  &7.32   &$\mathrm{C_{3V}}$($\mathrm{e}$/$\mathrm{e}$) &1.02/0.01    &2.44 &1.52  \\
&$\mathrm{O_SCu_{S}}^{-1}(\beta)$    &1.66 &1.09 &5.68  &$\mathrm{C_{3V}}$($\mathrm{e}$/$\mathrm{a_{1}}$)    &1.02/0.01  &2.44 &1.52\\
&$\mathrm{Ni_SMo_{S}}^{0}(\alpha)$   &0.91  &0.46    &1.74  &$\mathrm{C_{3V}}$($\mathrm{e}$/$\mathrm{a_{1}}$)  &7.39/0.01    &9.21 &7.37  \\
&$\mathrm{Mn_{Mo}Mn_{Mo}}^{0}(\beta)$    &1.78   &1.04   &6.78  &$\mathrm{C_{s}}$($\mathrm{a}$/$\mathrm{a}$)  &1.22/0.40     &1.70 &3.54         \\
&$\mathrm{O_{S}Ga_{Mo}}^{-1}(\alpha)$    &1.73  &1.10  &5.92 &$\mathrm{C_{s}}$($\mathrm{a}$/$\mathrm{a}$) &0.45/0.08       &3.23 &3.69   \\
\hline
\\[-1em]
GeS  &$\mathrm{Co_{Ge}Ge_{S}}^{1}(\beta)$ &3.12 &0.20 &9.65   &$\mathrm{C_{1}}$($\mathrm{a}$/$\mathrm{a}$)  &0.13/0.03 &3.41 &3.28\\
\hline
\\[-1em]
SiS$_2$ &$\mathrm{Co_{Si}}^{-1}(\beta)$  &1.80  &0.27  &3.95    &$\mathrm{C_{1}}$($\mathrm{a}$/$\mathrm{a}$) &7.31/1.66 &2.33 &2.49\\
&$\mathrm{O_SV_{Si}}^{-1}(\alpha)$  &2.02 &0.28  &4.19   &$\mathrm{C_{1}}$($\mathrm{a}$/$\mathrm{a}$) &-2.5/-0.28 &-1.76 &-1.68\\
&$\mathrm{O_SV_{Si}}^{-1}(\beta)$  &2.89  &3.03  &9.91   &$\mathrm{C_{1}}$($\mathrm{a}$/$\mathrm{a}$) &-2.5/-0.28  &-1.76 &-1.68\\
\hline
\\[-1em]
CaI$_2$  &$\mathrm{S_{Ca}Ca_{I}}^{1}(\alpha)$  &2.50 &3.91  &5.28   &$\mathrm{C_{1}}$($\mathrm{a}$/$\mathrm{a}$) &0.33/0.01 &3.94 &3.11\\
&$\mathrm{N_{Ca}Ca_{I}}^{0}(\alpha)$ &1.06 &3.96 &1.26   &$\mathrm{C_{1}}$($\mathrm{a}$/$\mathrm{a}$) &0.31/0.00 &5.0 &4.17\\
&$\mathrm{Cu_{Ca}Cu_{Ca}}^{0}(\alpha)$    &3.38 &3.35 &9.42  &$\mathrm{C_{1}}$($\mathrm{a}$/$\mathrm{a}$) &-0.30/-0.01 &6.95 &10.28\\
 \hline
 \\[-1em]
 MgBr$_2$   &$\mathrm{O_{Br}Fe_{Mg}}^{1}(\beta)$   &1.25 &0.42 &3.14  &$\mathrm{C_{1}}$($\mathrm{a}$/$\mathrm{a}$) &0.40/0.01 &4.48 &5.25\\
                 
    \hline \hline
    \end{tabular}
    \caption{\textbf{Properties of top candidate defects.} Zero-phonon lines (ZPL) (eV), momentum displacements ($\Delta$Q), Huang-Rhys factors (HR), zero field splitting (ZFS)(GHz) and formation energies $\mathrm{E^{f(A/B_{poor})}}$ (eV)  of all interesting defects.  Here, $\alpha$ refers to the majority spin channel and $\beta$ to the minority spin channel. The point group of the defect and the symmetry of HUMO and LUMO levels is mentioned (Pt. group(H, L)), which may help to identify the polarization of light emitted during emission process}
    \label{table:PL_results}
\end{table*}
 
\section{Conclusions}\label{sec:summaryoutlook}
With the aim of identifying novel emission centers with high-spin ground states and narrow photoluminescence (PL) line shapes, we used a workflow of first-principles calculations to screen more than 5000 point defects in ten different large band gap 2D materials. The most promising emitters were found in the host materials BN, MgI$_{2}$, MoS$_{2}$, MgBr$_{2}$, and CaI$_{2}$. Out of the ten materials considered, only MoS$_{2}$, GeS and SiS$_{2}$ host defects with spin coherence times larger than 1ms. While we did not find any sharp emitters in GeS and SiS$_{2}$, the triplet defects $\mathrm{v_SMo_S}^{0}$ and $\mathrm{Ni_SMo_{S}}^{0}$ in MoS$_{2}$ combine long coherence times with narrow emission spectra making them interesting for quantum technology applications.  We found that materials with higher in-plane stiffness are more likely to host point defects with low Huang-Rhys factors. All the calculated properties including atomic structures, formation energies, hyperfine coupling constants, zero-field splitting parameters and PL line shapes will be available in the open QPOD database\cite{bertoldo2022quantum}

\section{Methods}

\subsection{Density functional theory calculations}
All (spin-polarized) calculations are performed using the GPAW electronic structure code \cite{enkovaara2010electronic} with the PBE xc-functional \cite{perdew1996generalized}. 
We use a plane wave basis set with a cut off of $800$ eV and $k$-point density of 3 \AA for the structural relaxations and ground state calculations. We use the same Fermi smearing of $0.02$ eV for all calculations but for systems that are difficult to converge the Fermi-smearing is increased slightly. For the initial screening of triplets with loose computational parameters we use a single $k$-point ($\Gamma$-point only) and a plane wave cut off of $500$ eV. The supercell is kept fixed and atoms are fully relaxed until forces are below $0.01$ eV/\AA. We use the Pulay mixing scheme \cite{pulay1980convergence} where total density and magnetisation densities are treated separately. The excited states are calculated using the DO-MOM method\cite{ivanov2021method, levi2020variational}, where the maximum step length, $\rho_\mathrm{max}$, for the quasi-Newton search direction is chosen to be $0.2$. Both the phonons and excited states are calculated at the $\Gamma$-point. 

\section{Data availability}
The data generated will be made available in QPOD database. 

\section{Code availability}
The ASR recipe scripts used in the QPOD workflow are available at: \url{https://gitlab.com/asr-dev/asr/-/tree/defect\_formation/asr}.

\section*{References}
\bibliographystyle{iopart-num}
\bibliography{main_bib}

\section{Acknowledgments}
This work is supported by Novo Nordisk Foundation Challenge Programme 2021: Smart nanomaterials for applications in life-science, BIOMAG Grant No. NNF21OC0066526.  We acknowledge funding from the European Research Council (ERC) under the European Union’s Horizon 2020 research and innovation program Grant No. 773122 (LIMA) and Grant agreement No. 951786 (NOMAD CoE). K. S. T. is a Villum Investigator supported by VILLUM FONDEN (grant no. 37789). F.N has received funding from the European Union’s Horizon 2020 research and innovation program under the Marie Skłodowska-Curie grant agreement No. 899987. (EuroTechPostdoc2).

\section{Author contributions}
S.A. developed the initial concepts and discussed with all the authors. S.A. developed the workflow and performed analysis of the excited states, PL line shapes  and magneto-optical properties. F.N. developed the workflow and performed analysis of formation energies, host stiffness and uploaded the data to QPOD. S.M. contributed the analysis of CC-Diagram and symmetry related properties. F.B. developed the defect generation recipe. J.J.M. helped in implementation of the HF and ZFS tensor in GPAW. K.S.T. supervised the project and helped editing the manuscript. All authors discussed and modified the manuscript.  

\section{Competing interests}
The authors declare no competing interests.

\end{document}